\documentclass{aa}
\usepackage[varg]{txfonts}
\usepackage{graphicx}
\usepackage{dblfloatfix}
\usepackage{epstopdf}
\usepackage{pdflscape}
\usepackage{booktabs}
\usepackage{txfonts}
\usepackage{gensymb}
\usepackage{color}
\usepackage{url}
\usepackage{multirow}
\usepackage{amsmath}
\usepackage{amstext}
\usepackage{natbib}
\usepackage{longtable}
\usepackage{float}
\usepackage[export]{adjustbox}

\usepackage{caption}
\usepackage{siunitx} 
\usepackage{threeparttablex} 

\begin{document}

\title{Coma environment of comet C/2017 K2 around the water ice sublimation boundary observed with VLT/MUSE}

\author{Yuna G. Kwon\inst{\ref{inst1}\thanks{Based on observations made with ESO Telescopes at the Paranal
Observatory under program 109.24F3.001 (PI: Y. G. Kwon)}}~\and Cyrielle Opitom\inst{\ref{inst2}}~\and Manuela Lippi\inst{\ref{inst1},\ref{inst3}}}

\institute{Institut f{\" u}r Geophysik und Extraterrestrische Physik, Technische Universit{\" a}t Braunschweig,  Mendelssohnstr. 3, 38106 Braunschweig, Germany (\email{y.kwon@tu-braunschweig.de}\label{inst1})
\and Institute for Astronomy, University of Edinburgh, Royal Observatory, Edinburgh, EH9 3HJ, UK\label{inst2}
\and Current affiliation: INAF - Osservatorio Astrofisico di Arcetri - Largo Enrico Fermi, 5, 50125 Firenze, Italy\label{inst3}}

\date{Received \today / Accepted ---}

\abstract {We report a new imaging spectroscopic observation of Oort-cloud comet C/2017 K2 (hereafter K2) on its way to perihelion at 2.53 au, around a heliocentric distance where H$_{\rm 2}$O ice begins to play a key role in comet activation. Normalized reflectances over 6 500--8 500 \AA\ for its inner (cometocentric distance $\rho$ $\approx$ 10$^{\rm 3}$ km) and outer ($\rho$ $\approx$ 2 $\times$ 10$^{\rm 4}$ km) comae are 9.7$\pm$0.5 and 7.2$\pm$0.3 \% (10$^{\rm 3}$ $\AA$)$^{\rm -1}$, respectively, the latter being consistent with the slope observed when the comet was beyond the orbit of Saturn. The dust coma of K2 at the time of observation appears to contain three distinct populations: mm-sized chunks prevailing at $\rho$ $\lesssim$ 10$^{\rm 3}$ km; a 10$^{\rm 5}$-km steady-state dust envelope; and fresh anti-sunward jet particles. The dust chunks dominate the continuum signal and are distributed over a similar radial distance scale as the coma region with redder dust than nearby. 
They also appear to be co-spatial with OI$^{\rm 1}$D, suggesting that the chunks may accommodate H$_{\rm 2}$O ice with a fraction ($\gtrsim$1 \%) of refractory materials. The jet particles do not colocate with any gas species detected. The outer coma spectrum contains three significant emissions from C$_{\rm 2}$(0,0) Swan band, OI$^{\rm 1}$D, and CN(1,0) red band, with an overall deficiency in NH$_{\rm 2}$. Assuming that all OI$^{\rm 1}$D flux results from H$_{\rm 2}$O dissociation, we compute an upper limit on the water production rate Q$_{\rm H_{\rm 2}O}$ of $\sim$7 $\times$ 10$^{\rm 28}$ molec s$^{\rm -1}$ (with an uncertainty of a factor of two). The production ratio $\log$[Q$_{\rm C_{\rm 2}}$/Q$_{\rm CN}$] of K2 suggests that the comet has typical carbon-chain composition, with the value potentially changing with increasing distance from the Sun. Our observations suggest that water ice-containing dust chunks ($>$0.1 mm) near K2's nucleus emitted beyond 4 au may be responsible for its very low gas rotational temperature and the discrepancy between its optical and infrared lights reported at similar heliocentric distances.
}


\titlerunning{The dust and gas coma of C/2017 K2}

\authorrunning{Y. G. Kwon et al.}

\maketitle

\section{Introduction \label{sec:intro}}

Comets, one of the least-altered planetesimals in our solar system, consist of dust and ice. Sublimation of ice entrains dust and gas molecules, forming a cometary coma whose development is primarily determined by the nature of its nucleus. As such, identifying activity provides valuable insight into the characteristics of the building blocks that make up comet nuclei.

Comet C/2017 K2 (PANSTARRS) (hereafter K2) is an active Oort-cloud comet crossing heliocentric distances of $\sim$2.5--3.0 au from the Sun in July 2022 where the main trigger of comet outgassing transitions from supervolatile ices to H$_{\rm 2}$O ice (so-called water ice sublimation boundary;  \citealt{Blum2014,Womack2017,Gundlach2020}). K2 already developed a 10$^{\rm 5}$-km-scale coma outside Saturn \citep{Jewitt2017} that would be initiated in the Kuiper Belt \citep{Jewitt2021}. With a dust production rate up to $\sim$10$^{\rm 3}$ times that of most comets at similar distances from the Sun (e.g. \citealt{Garcia2020}), K2 offers a unique opportunity to study an activity regime previously not observed in detail. 

\begin{figure}[!b]
\centering
\includegraphics[width=8.5cm]{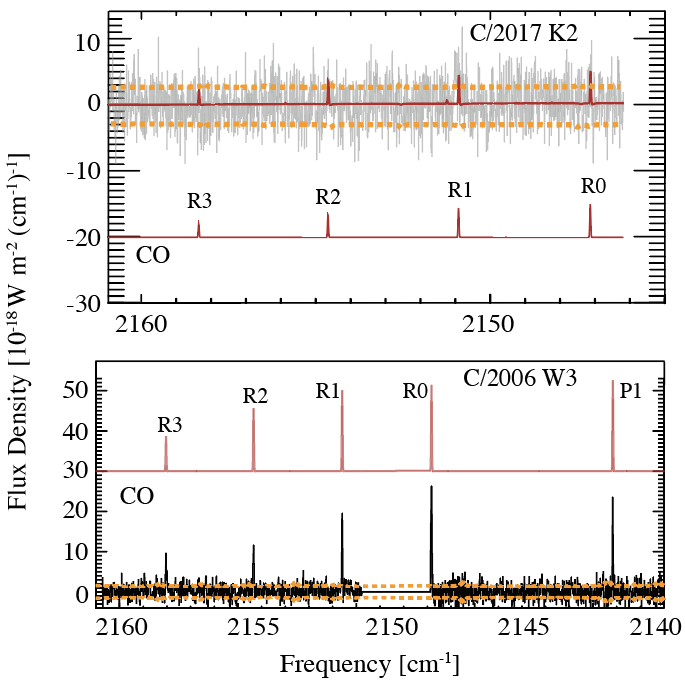}
\caption{Comparison between CO spectra of C/2017 K2 observed in July with ESO/CRIRES$^+$ (top panel, Lippi et al., in preparation) and C/2006 W3 (Christensen), observed in October 2009 with ESO/CRIRES (bottom panel, adapted from \citealt{Bonev2017}). The comets were at heliocentric distances of 2.75 au and 3.25 au, respectively. In both panels, the red lines indicate the CO model, while the dotted yellow lines are the $\pm 1\sigma$ noise. C/2006 W3 (Christensen) shows very strong CO emission lines even if at a larger heliocentric distance than C/2017 K2, while CO lines for the latter are at the noise level.}
\label{k2co}
\end{figure}

During monitoring K2 as it crossed over the water ice sublimation boundary, observations with ESO/CRIRES$+$ at the beginning of July revealed an unexpected discrepancy at about 2.7 au: despite the brightness and expected activity of the comet, the infrared emission features of gas molecules and dust continuum are barely discernible except for a few faint lines of CO (Fig. \ref{k2co}, \citealt{Lippi2022}, Lippi et al., in prep.), CH$_{\rm 4}$, and C$_{\rm 2}$H$_{\rm 6}$  compared to what is expected from their optical counterparts \citep{Jehin2022}. For comets in general, the infrared domain over $\sim$3--5 $\mu$m contains fundamental fluorescence emissions from molecules that are released directly from the nucleus (i.e., ``primary or parent molecules'', e.g. H$_{\rm 2}$O, CH$_{\rm 4}$, CO, and CH$_{\rm 3}$OH; \citealt{Bockelee-Morvan2004}), superimposed on a combination of solar-reflected light and thermal dust continuum. As the molecules move outward from the nucleus, they interact with solar photons and become photo-dissociated. The photo-dissociated products (i.e., ``secondary or daughter molecules'', e.g. C$_{\rm 2}$, NH$_{\rm 2}$, and CN; \citealt{Feldman2004}) produce emission features in the optical domain over $\sim$0.4--1.0 $\mu$m. Bright emissions from daughter molecules usually indicate the presence of corresponding parent molecules in the infrared \citep{Mumma2011,Biver2022}: for example, a high amount of OH measured in the optical, should correspond to a similar amount of its parent molecule H$_2$O sampled in the infrared. In this regard, the observed discrepancy in K2 would suggest an alternative source for the observed daughter molecules rather than ice embedded in the nucleus or some physical processes that we are still unaware of. Moreover, the very faint thermal dust continuum suggests a low-temperature environment and dust with unusual size distributions and/or compositions.

The lack of spatially resolved information regarding what happens to comets farther from the Sun makes it difficult to determine whether the observed discrepancy is typical of comets crossing the water ice sublimation boundary or if it is unique to comet K2. Hence, we conducted one-epoch observation using MUSE, the Multi Unit Spectroscopic Explorer, at 2.53 au where the comet is still within the same activity regime as the above-mentioned optical and infrared observations. With MUSE offering simultaneous spectral and spatial information on dust and gas coma species, we aim to investigate how those species compose coma signals and therefrom suggest possible explanations for the observed discrepancy.

\begin{table*}[!t]
\centering
\caption{Geometry and instrument settings of the observations of C/2017 K2 (PANSTARRS) on UT 2022 July 29}
\begin{tabular}{c|c|c|c|c|ccccc|c}
\hline
\hline
Telescope/ &  \multirow{2}{*}{Median UT} & \multirow{2}{*}{$N$} & {Exptime} & \multirow{2}{*}{Airmass} & $m_{\rm V}$ & $r_{\rm H}$ & $\Delta$ & $\alpha$ & $\nu$ & \multirow{2}{*}{Std. star} \\
Instrument & & & (sec) &  & (mag) & (au) & (au) & (\degree) & (\degree) & \\
 \hline
 \hline
VLT-UT4/ & 02:25:39 & \multirow{2}{*}{7} & \multirow{2}{*}{400} & 1.16 & \multirow{2}{*}{12.549} & \multirow{2}{*}{2.530} & \multirow{2}{*}{1.850} & \multirow{2}{*}{20.1} & \multirow{2}{*}{294.9} & \multirow{2}{*}{GJ 754}\\
MUSE & (01:40:33 -- 03:10:44) & &  & (1.06--1.26) & & & & & & \\
 \hline
 \hline
\end{tabular}
\tablefoot{Top headers: $N$, number of exposures; Exptime, total integration time of each cube in seconds; $m_{\rm V}$, apparent $V$-band magnitude provided by the JPL Horizons (http://ssd.jpl.nasa.gov/?horizons); $r_{\rm H}$ and $\Delta$, median heliocentric and geocentric distances in au, respectively; $\alpha$, median phase angle (angle of Sun--comet--observer) in degrees; $\nu$, median true anomaly in degrees. Numbers in parentheses show ranges of UT and airmass during the observation.} 
\label{t01}
\end{table*}

\section{Observations and Data Analysis\label{sec:obsdata}}

This study is based on single-epoch data carried out under the program (ID: 109.24F3.001) of the Director's Discretionary Time of the European Southern Observatory (ESO).

\subsection{Observations\label{sec:obs}}

MUSE is an Integral Field Unit (IFU) spectrograph mounted on the UT4 telescope of the Very Large Telescope (VLT) at the Paranal Observatory (70\degree24\arcmin10\farcs1W, 24\degree37\arcmin31\farcs5S, 2 635 m) in Chile \citep{Bacon2010}. MUSE in Wide Field Mode (WFM)\footnote{\url{https://www.eso.org/sci/facilities/paranal/instruments/muse/doc/ESO-261650_MUSE_User_Manual.pdf}} splits a field of view (FoV) of 1$\arcmin$ $\times$ 1$\arcmin$ into 24 channels that are further sliced into 48 slitlets each sampling 15$\arcsec$ $\times$ 0\farcs2 without Adaptic Optics (WFM-noAO mode) with a pixel resolution of 0\farcs2. Each slice offers a medium-resolution (1.25 $\AA$ interval) spectrum over the nominal wavelength coverage of 4 800--9 300 $\AA$. 
Separate sky observations were taken at positions $\gtrsim$550\arcsec\ from the nucleus for every two target observations in the order of O-S-O-O-S-O, where O and S denote object and sky exposure respectively. A combination of sky acquisition and $+$90\degree\ position angle rotations averages out optics-induced signals, rejects cosmic rays, and thus improves the signal-to-noise ratio. We conducted a 1.5-hour observation for K2 on UT 2022 July 29. The atmospheric seeing ranged from $\sim$0\farcs4 to 1\farcs2, with the maximum being an instant peak otherwise seeing remained at $\sim$0\farcs5 most of the time. A journal of observations is given in Table \ref{t01}. 

\subsection{Data analysis\label{sec:data}}

The basic data reduction (i.e., bias and sky subtraction, flat-fielding, wavelength and flux calibration, and telluric absorption correction) and construction of 3D cubes were all conducted via the ESO/MUSE Data Reduction Pipeline (DRP; \citealt{Weilbacher2020}). A standard star observed on the same night as our observation was used for sky subtraction and telluric correction. Since DRP corrects first-order telluric lines feature to satisfactory levels (e.g. the strongest O$_{\rm 2}$ telluric line at $\sim$7 600 $\AA$ is well canceled out in Fig. \ref{Fig01}a), we did not apply further modeling. 
The resulting seven cubes (the 8$^{\rm th}$ cube was discarded due to its large airmass and incomplete FoV coverage) showed no rotational variation in the coma morphology. We therefore median combined them with a 5$\sigma$-clipping using the open-source software, MUSE Python Data Analysis Framework (\texttt{mpdaf})\footnote{\url{https://mpdaf.readthedocs.io/en/stable/index.html}}. In the following, all results are based on the quantities derived from the median-combined cube.

A different methodology was applied for the analysis of the oxygen lines described in Sect. \ref{sec:res3}. Indeed, oxygen-forbidden lines at 5 577, 6 300, and 6 364 $\AA$ are also present in the sky. At the spectral resolution of MUSE, we cannot resolve the telluric and cometary lines. In order to exploit the information contained in those lines, we have run a separate data reduction without applying any sky subtraction. For this analysis, all 8 data cubes are considered separately.

\section{Results\label{sec:res}}

This section examines the dust and gas coma environments in K2. The general circumstances are introduced, followed by descriptions of each coma component.

\subsection{General outlines \label{sec:res1}}

\begin{figure}[!t]
\centering
\includegraphics[width=8.5cm]{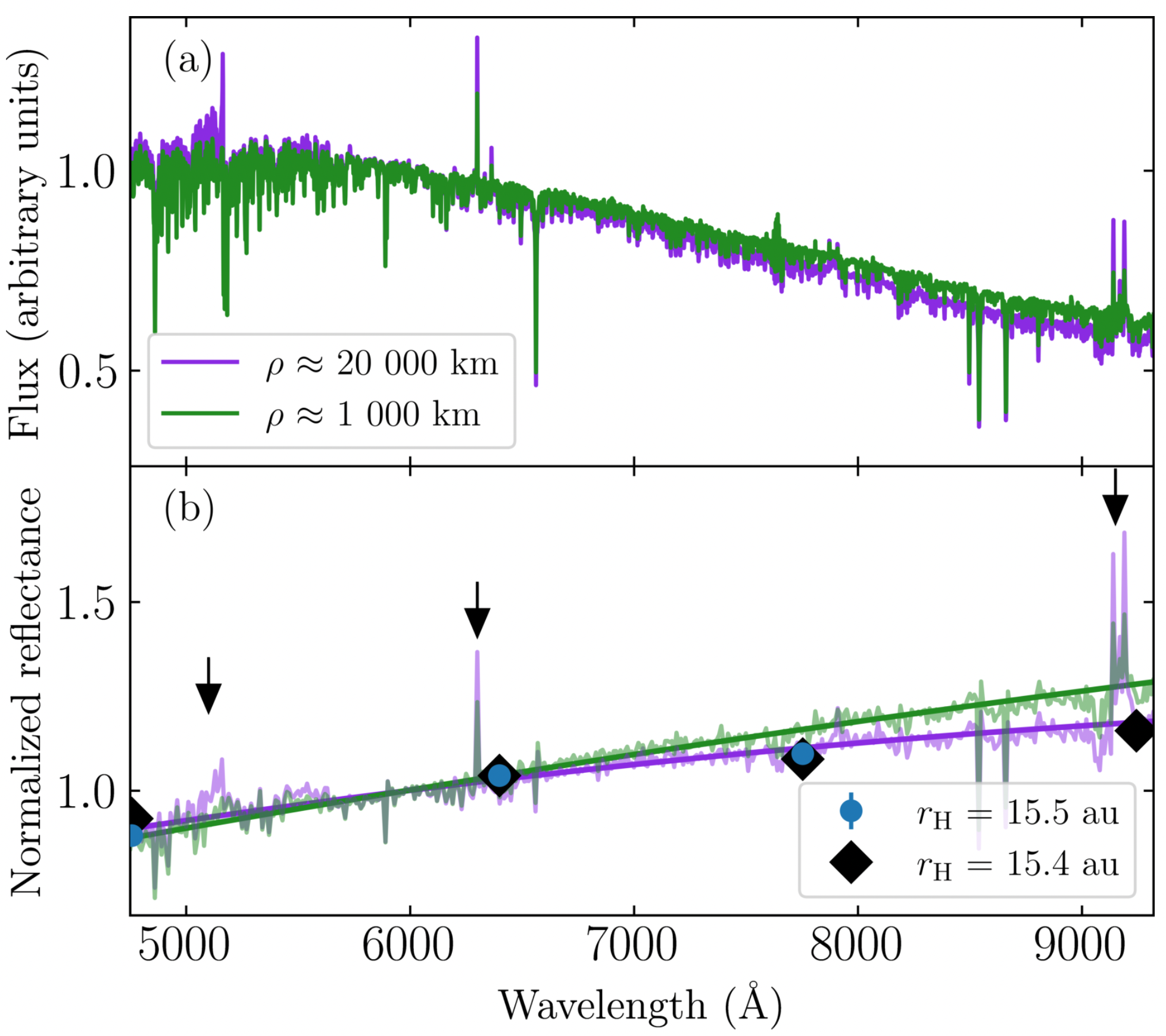}
\caption{(a) Unbinned spectrum of the K2 coma (with a step of 1.25 $\AA$) extracted over smaller (corresponding to the cometocentic distance $\rho$ of $\approx$10$^{\rm 3}$ km; green) and larger ($\rho$ $\approx$ 2 $\times$ 10$^{\rm 4}$ km; purple) radius circular apertures. (b) Relative reflectance of K2 normalized at 6 000 $\AA$. K2's spectra are binned to match the spectral resolution (5 $\AA$) of the solar spectrum. Three arrows mark significant emission features of gas species: C$_{\rm 2}$(0,0) band at $\sim$5 100 $\AA$, OI$^{\rm 1}$D line at 6 300 $\AA$, and CN(1,0) red band at $\sim$9 200 $\AA$. The circles and diamonds are the data of K2 taken at $r_{\rm H}$ $\sim$15.5 and $\sim$15.4 au, respectively \citep{Meech2017}.}
\label{Fig01}
\end{figure}

Figure \ref{Fig01} shows spectra of K2 extracted over two different aperture radii, each corresponding to the cometocentric distance $\rho$ $\approx$ 10$^{\rm 3}$ km (inner coma, green) and 2 $\times$ 10$^{\rm 4}$ km (outer coma, purple). 
OI$^{\rm 1}$D $^{\rm 1}$D--$^{\rm 3}$P line at 6 300 $\AA$ and CN A$^{\rm 2}\Pi$--X$^{\rm 2}\Sigma^{\rm +}$ (1,0) red band at $\sim$9 200 $\AA$ are evident in both the inner and outer coma regions, while C$_{\rm 2}$ d$^{\rm 3}\Pi_{\rm g}$--a$^{\rm 3}\Pi_{\rm u}$ (0,0) Swan band at $\sim$5 100 $\AA$ is discernible only in the outer part, possibly due to dust contamination.  
NH$_{\rm 2}$ $\tilde{\rm A}^{\rm 2}$A$_{\rm 1}$--$\tilde{\rm X}^{\rm 2}$B$_{\rm 1}$ band signals that are expected to be present over 5 600--7 400 $\AA$ are negligible as a whole.

The extracted spectra were then divided by the reference solar spectrum taken at an airmass of 1.5 provided by the American Society for Testing and Material\footnote{\url{https://www.nrel.gov/grid/solar-resource/spectra-am1.5.html}} and binned to match the spectral resolution (5 \AA) of the solar spectrum. In spite of a slight mismatch in airmass between the two spectra, the resultant reflectance (Fig. \ref{Fig01}b) provides a useful measurement of the K2's gross dust color. 
Dust in both coma regions is redder than the Sun, slightly concaving down at longer wavelengths. The inner coma is redder than the outer coma: fitted linearly over 6 500--8 500 \AA, the normalized reflectivity $S^{'}$ \citep{Jewitt1986} is 9.7$\pm$0.5 and 7.2$\pm$0.3 \% (10$^{\rm 3}$ $\AA$)$^{\rm -1}$ for the inner and outer comae, respectively. Both $S^{'}$ are slightly bluer than, for instance, 2I/Borisov \citep{Bannister2020} and 67P/Churyumov-Gerasimenko \citep{Snodgrass2016} at similar heliocentric distances in their inbound orbits but well in the average range of solar system comets \citep{Solontoi2012}. The outer coma slope is consistent with that obtained at 15.5 and 15.4 au from the Sun (diamonds and circles covering $\rho$ $\sim$ 56 000 km in Fig. \ref{Fig01}b;  \citealt{Meech2017}), indicating that the large-scale coma has remained relatively constant.

\subsection{Dust coma environment: three distinct populations \label{sec:res2}}

\begin{figure}[!t]
\centering
\includegraphics[width=9cm]{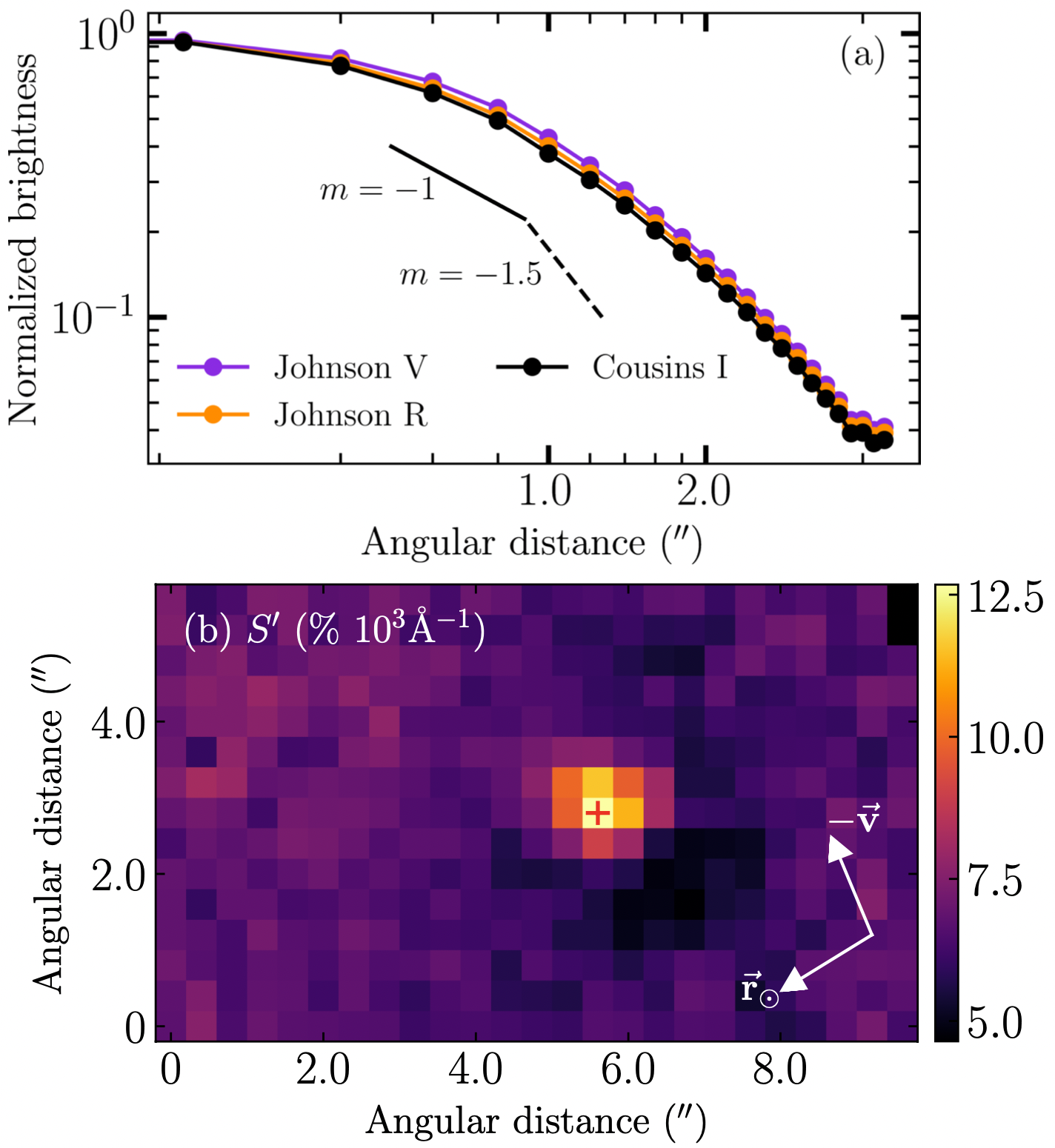}
\caption{(a) Azimuthally-averaged radial profiles of K2 in the Johnson V and R and Cousins I filters. Two guidelines with the slope of $-$1 and $-$1.5 are plotted for comparison. (b) Map of the normalized reflectivity $S^{'}$. A 2 $\times$ 2 binning is employed such that each pixel covers 0\farcs4. The red plus ($+$)  marks the optocenter. The negative velocity  ($-$${\bf{\vec{v}}}$) and anti-sunward (${\bf{\vec{r_{\rm \odot}}}}$) vectors are given. North is up, east to the left.}
\label{Fig02}
\end{figure}

We first create coma images of K2 by integrating the flux bracketed by three broadband filters V, R, and I whose azimuthally-averaged radial profile is shown in Figure \ref{Fig02}a. The expected slope ranges for steady-state coma ($-$1 and $-$1.5; \citealt{Jewitt1987}) are plotted for comparison.
The profiles are similar to one another, showing a shallower slope than $-$1 from the center out to $\sim$4 pixels (corresponding to $\rho$ $\sim$ 10$^{\rm 3}$ km) followed by a steady decrease\footnote{The central part should be interpreted with caution due to possible seeing effects. We confirm that all coma structures in the median-combined cube discussed in this paper are consistently present in single cubes obtained under optimal seeing conditions of 0\farcs39 and 0\farcs5. To be conservative, only features spanning $\gtrsim$0\farcs8 visible in every single cube are considered valid.}.
In a 2 $\times$ 2-binned $S^{'}$ distribution (Fig. \ref{Fig02}b), this central region of shallow radial profile corresponds to the coma region on the scale of $\rho$ $\approx$ 10$^{\rm 3}$ km having more reddish colors than the outer coma, with a lack of large-scale asymmetry. The flux originating from this reddened region is distributed on a radial scale similar to the bright central areas in continuum and gas-emitting signals (Fig. \ref{Figap01}a). We refrain from making a definitive statement regarding the alignment of this feature since the binned image here shows its slight sunward extension, but not the original image.
The red dust color accompanied by continuum enhancement suggests that the dust clusters in this near-nucleus area are possibly dark (containing abundant carbonaceous materials like typical cometary dust; \citealt{Levasseur-Regourd2018,Filaccione2020}), consolidated, large chunks that are less sensitive to solar radiation pressure \citep{Hadamcik2009,Kwon2022a}. 
We will discuss this implication in Sect. \ref{sec:dis}. 

\begin{figure*}[!h]
\centering
\includegraphics[width=\textwidth]{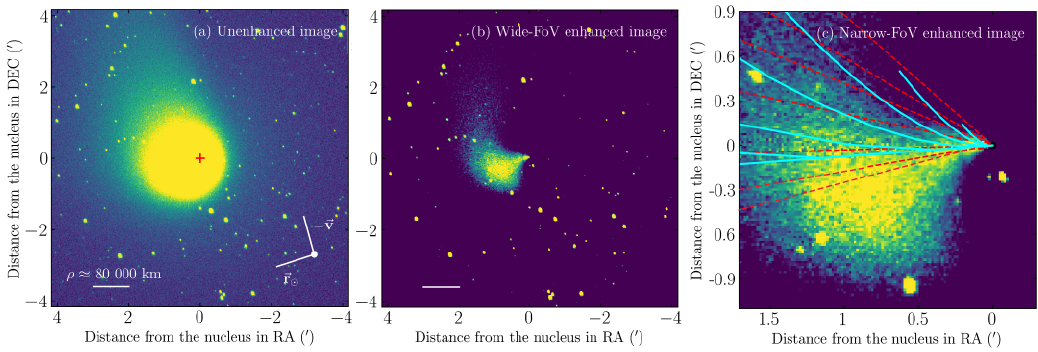}
\caption{(a) Unenhanced ZTF zg-band image of K2 on UT 2022 July 26, with the vector notation used in Figure \ref{Fig02}. The red plus ($+$) marks the position of the nucleus as defined by the peak flux. (b) zg-band image enhanced by dividing out of an azimuthal median profile \citep{Samarasinha2013}. (c) Synchrones and syndynes of the K2 coma. The red dashed lines are synchrones, indicating the locations of dust ejected at 1, 5, 15, 30, 60, 90, 120, and 180 days prior to the ZTF observation from bottom to top. The cyan curves are syndynes, where each has a constant $\beta$ of 1, 0.3, 0.1, 0.03, 0.01, 0.003, and 0.001 clockwise from the bottom. A nonzero initial outflow velocity is required to explain dust spreading from south to southeast which is not explained by the combination of dust parameters tested.
Brightness levels were arbitrarily adjusted on a linear scale in order to highlight the coma feature of interest in each image.
}
\label{Fig03}
\end{figure*}

Next, we investigate any asymmetric structure embedded in the outer envelope by dividing out an azimuthal median profile from a broadband image  \citep{Samarasinha2013}. To this end, we utilize a zg-band image of K2 from the Zwicky Transient Facility\footnote{\url{https://irsa.ipac.caltech.edu/applications/ztf/}} (ZTF) archive \citep{Masci2019} taken on the nearest date ($r_{\rm H}$ $\sim$ 2.55 au on UT 2022 July 26) from our observation, as the MUSE's FoV is not wide enough to examine macroscopic coma structures. Figure \ref{Fig03}a presents the ZTF image before enhancement. At first glance, the tail edge of the comet appears sharp along its negative velocity direction. This feature is indicative of the protracted ejection of large dust particles (their large inertia anchors them to the comet position from which they were released; e.g. \citealt{Ishiguro2007}), which is more pronounced in amateur images taken near perihelion\footnote{\url{https://cometografia.es/2017k2-panstarrs-2022-09-18/#more-15037}}. Figure \ref{Fig03}b presents the enhanced image showing a conspicuous anti-sunward jet that curves into the negative velocity direction at its outer end. Its lack of spatial correlation with gas species detected (Sect. \ref{sec:res3}) suggests that this feature is probably dusty.
We replicate this enhanced feature of K2 using a Finson-Probstein mechanism \citep{Finson1968a,Burns1979} assuming zero ejection velocity where inward solar gravity forces ($F_{\rm grav}$) and outward solar radiation pressure ($F_{\rm rad}$) determine dust trajectory in the outer coma. Using Eq. 1 and constants in \citet{Kwon2022c}, we calculate a parameter $\beta$ (a ratio of $F_{\rm rad}$ to $F_{\rm grav}$) that demonstrates the relative importance of solar radiation pressure on dust and is directly related to dust density and size. We specify each coma location with a combination of a synchrone (each represents dust with different $\beta$ but ejected at the same epoch) and syndyne (each represents dust with the same $\beta$ but ejected at a different epoch) so as to compare their distribution with the observed coma morphology. For this modeling, we consider a range of ejection epochs from $\sim$180 days ($>$4.1 au, well outside of the water ice sublimation boundary) to one day prior to our observation, while testing dust with $\beta$ from 1 down to 10$^{\rm -4}$. Additional forces that can alter dust motions in the coma (e.g. gas drag force, sublimation of ice embedded, and dust fragmentation) are not considered here.

Figure \ref{Fig03}c shows the synchrones and syndynes plotted over the enhanced image, along with the ejection times and values considered. Dust with $\beta$ of 10$^{\rm -4}$ (the smallest value tested, of the order of millimeters) is not plotted here because its extension is too short to be distinguished from the nucleus position on the given FoV. Only dust ejected recently ($<$5 days, at best $\sim$1 day prior to the observation) and with high mobility ($\beta$ $\gtrsim$ 0.3) is capable of reproducing the jet's position angle. However, the southern--southeastern distribution of the anti-sunward dust spread that even sufficiently small particles ($\beta$ = 1, of the order of 0.1 $\mu$m) cannot reproduce, strongly indicates that our simple assumption of zero ejection velocity does not hold and thus requires invoking non-zero initial speeds, such as rocket force from sublimating ice particles \citep{Kelley2013}. 
Nevertheless, given that there is no evidence of dust being redistributed immediately into the anti-sunward direction by solar radiation pressure with the end of the jet curving into the negative velocity direction, dust composing this jet feature may share similar properties as dust on a tail along the negative velocity direction (i.e., an order of millimeters). We confirmed the presence of a similar feature in the zg- and zr-band ZTF images taken on UT 2022 August 15.

\subsection{Gas coma environment\label{sec:res3}}

\subsubsection{OI$^{\rm 1}$D forbidden lines}

Forbidden oxygen lines at 5 577.339 $\AA$, 6 300 $\AA$, and  6 363.776 $\AA$ have been observed in a large number of comets, usually using high-resolution spectrographs. Since we cannot separate the cometary and telluric contributions at the spectral resolution of MUSE, we used a combination of spatial and spectral information to exploit the information contained in these oxygen lines. Following the methodology described in \citet{Opitom2020}, we used datacubes for which the sky was not subtracted to create maps of the spatial distribution of the (dust-subtracted) flux in the three oxygen lines (Fig. \ref{Figap04}). On these maps, you can see the contribution of the comet as a peaked contribution decreasing relatively rapidly, on top of a constant contribution from the sky. To produce the maps of the 5 577.339 and 6 363.776 $\AA$ lines, we re-centered and co-added the individual maps from 7 datacubes. The sky contribution was then measured and subtracted using an annulus of 2\arcsec\ width at 20\arcsec\ from the comets. For the 6 300 $\AA$ line, we noticed that the emission was filling the FoV in some cases. We thus only combined 4 maps, for which the comet was offset from the center of the field (due to centering issues during the observations) and for which the sky contribution could be measured at the edge of the image. For this case, the sky was measured from a small 1\arcsec\ aperture at the lower right edge of the field, which was part of the map free from comet emission. 

We measured the intensity ratio of the green line to the sum of the two red lines (G/R) in two circular apertures for all maps and obtain G/R = 0.17$\pm$0.02 in a 1\arcsec\ radius aperture and G/R = 0.13$\pm$0.02 in a 2\arcsec\ radius aperture. This technique can only produce an estimate of the G/R ratio, as at the spectral resolution of MUSE, we cannot exclude contamination from C$_2$ in our measurement of the 5 577.339 $\AA$ line. However, in the case of these observations of C/2017 K2, the C$_2$ emission from the (1,0) band is faint and only detected when the extraction is done with a sufficiently large aperture. Our measurements are consistent with what is measured at similar heliocentric distances by \cite{Decock2013} and with the effect of quenching that leads to higher G/R ratios in small apertures \citep{Decock2015}. Given that G/R of $\sim$0.1 is expected in purely H$_{\rm 2}$O-driven activity with a larger value (up to $\sim$0.35) in CO- and CO$_{\rm 2}$-dominating comet activity \citep{Decock2013,Decock2015}, the result indicates that H$_{\rm 2}$O may play a leading role at our observing epoch, though some CO or CO$_{\rm 2}$ contributions cannot be ruled out.

Making the hypothesis that the OI$^{\rm 1}$D line at 6 300 \AA\ is only produced by the dissociation of H$_2$O, we can obtain a crude estimate of the water production rate. To do that, we measured the flux of the 6 300 \AA\ over a rectangular aperture (illustrated in Fig. \ref{Figap04}). Examination of the radial profile shows that within that aperture, the OI$^{\rm 1}$D flux decreases down to the background level at the edge of the field and we are sampling all the flux from the object in that quadrant. Since this aperture only covers 1/4 of the coma, we multiplied the measured flux by 4. To derive an estimate of the water production rate, we followed the procedure described in \citet{Opitom2020} and used branching ratios for the quiet Sun from \citet{Morgenthaler2001}. Assuming that we measured the full flux of the coma, we obtain a water production rate Q$_{\rm H_{\rm 2}O}$ from the OI$^{\rm 1}$D 6 300 \AA\ flux of 7 $\times$ 10$^{\rm 28}$ molec s$^{\rm -1}$, with an uncertainty of factor two, mainly due to the difficulty to subtract the sky component for this very extended emission. Our Q$_{\rm H_{\rm 2}O}$ estimate is consistent with those of active Oort-cloud comets at $\sim$2.5 au preperihelion distance (e.g. \citealt{Fink2009,Combi2018}).

\subsubsection{CN(1,0) and C$_{\rm 2}$(0,0) bands}

\begin{figure}[!b]
\centering
\includegraphics[width=8.9cm]{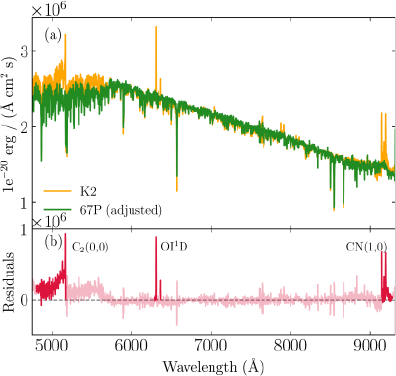}
\caption{(a) Unbinned spectra of K2 (orange) and 67P/Churyumov-Gerasimenko (green, quoted from \citealt{Opitom2020}) whose flux and slope are adjusted to those of K2. (b) Flux residuals after subtracting the adjusted 67P spectrum from the K2. Four emission features are highlighted by thick red lines whose flux and underlying continuum images are given in Figure \ref{Figap03}. Although C$_{\rm 2}$(1,0) at $\sim$5 500 $\AA$ is also marginally discernible, we did not create a flux map for it in this study due to its large uncertainty following continuum subtraction.}
\label{Fig04}
\end{figure}

The signals of the CN(1,0) and C$_{\rm 2}$(0,0) bands are separated by subtracting the continuum from the spectrum observed. 
Following \citet{Opitom2020}, we employed a reference dust spectrum of comet 67P/Churyumov-Gerasimenko taken by the same instrument (MUSE), adjusted its flux and slope to match with those of K2, and subtracted the reference from the K2 spectrum (the outer coma spectrum; purple in Fig. \ref{Fig01}a). Figure \ref{Fig04}a illustrates the adjusted reference spectrum to be subtracted. 
The subtraction was carried out for each pixel to produce maps of the spatial distributions of gas components.

\begin{figure}[!t]
\centering
\includegraphics[width=8.5cm]{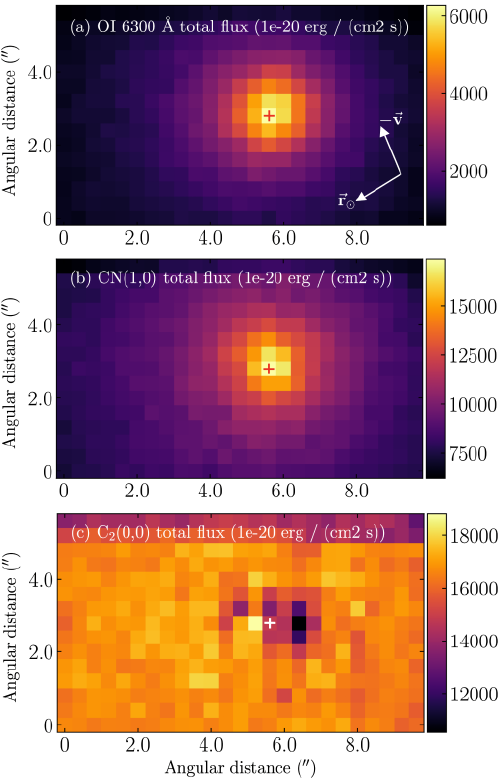}
\caption{2 $\times$ 2-binned flux distribution of (a) OI$^{\rm 1}$D line, (b) CN(1,0) red band, and (c) C$_{\rm 2}$(0,0) Swan band in units of 10$^{\rm -20}$ erg cm$^{\rm -2}$ s$^{\rm -1}$. The plus ($+$) marks the optocenter. The images have the same FoV, pixel scales, and alignment directions as those in Figure \ref{Fig02}b.
}
\label{Fig05}
\end{figure}

This process leaves two significant -- C$_{\rm 2}$(0,0) Swan band, and CN(1,0) red band -- and one fainter OI$^{\rm 1}$D 6 364 \AA\ emission signals (Fig. \ref{Fig04}b).
A lack of NH$_{\rm 2}$ bands, particularly NH$_{\rm 2}$ band which overlaps OI$^{\rm 1}$D at $\sim$6 300 \AA\ \citep{Fink1994}, attributes the observed feature to the latter (OI$^{\rm 1}$D also appears clearly in the image without removing the sky, supporting its cometary origin; Fig. \ref{Figap04}).
We integrated over the wavelengths highlighted in thick red in Figure \ref{Fig04}b and Figure \ref{Fig05} displays resultant integrated flux distributions for the three dominant gas species. 
Table \ref{t02} provides information on areas of integration, their flux, and production rates, while their ancillary information (unbinned flux distributions and underlying continuum distribution) can be found in Figure \ref{Figap03} alongside the information of OI$^{\rm 1}$D 6 364 \AA\ line. 
Both OI$^{\rm 1}$D and CN(1,0) are the strongest at the optocenter (`$+$' symbols in Fig. \ref{Fig05}) with slight sunward extension as in the distribution of the spectral reddening (Fig. \ref{Fig02}b). Meanwhile, C$_{\rm 2}$(0,0) flux peaks consistently off from the optocenter by $\sim$500--1 000 km, which could be due in part to an over-subtraction of dust near the optocenter.

\begin{table}[!t]
\centering
\small
\caption{Areas of integration and results of spectral analysis.}
\begin{tabular}{c|c|cc}
\hline
\hline
Gas&  Integration$^{\rm \star}$ & $F_{\rm X}$ & $Q_{\rm X}^{\dagger}$ \\
species & (\AA) &  ($\times$10$^{\rm -12}$ erg~cm$^{\rm -2}$~s$^{\rm -1}$) & ($\times$10$^{\rm 26}$ molec~s$^{\rm -1}$) \\
 \hline
 \hline
CN(1,0) & 9 141--9 269 & 2.0 $\pm$ 0.2 & 0.7 $\pm$ 0.1 \\
C$_{\rm 2}$(0,0) & 4 801--5 170 & 9.4 $\pm$ 1.0 & 1.1 $\pm$ 0.2 \\
 \hline
 \hline
\end{tabular}
\tablefoot{$^{\rm \star}$Integrated wavelength ranges. We consider integration areas slightly narrower than each gas species' theoretical range to minimize uncertainties from continuum subtraction. $^\dagger$We placed an arbitrary error of 20 \% taking into account the order-of-magnitude estimation of production rates.} 
\label{t02}
\end{table}

The integrated fluxes $F$ of C$_{\rm 2}$(0,0) and CN(1,0) bands over the aperture radius of 15\arcsec\ (Table \ref{t02}) are converted to their production rates $Q$ using the Haser model \citep{Haser1957}. We used conventional parameters of the model to provide $Q$ estimates on an order-of-magnitude basis. The outflow velocity was scaled by the inverse scaling law that $v$ = 0.58 $\times$ $r_{\rm H}^{\rm -2}$, where $r_{\rm H}$ is the heliocentric distance (2.53 au) \citep{Delsemme1982}. $g$-factors (fluorescence efficiencies) and daughter scalelengths for CN(1,0) and C$_{\rm 2}$ are from \citet{Lara2004} and \citet{A'Hearn1995}, respectively, and scaled as $r_{\rm H}$$^{\rm -2}$ (for the $g$-factors) and as $r_{\rm H}$$^{\rm 2}$ (for the scalelengths). Together with the Haser correction factors provided by \citet{Schleicher2010}\footnote{\url{https://asteroid.lowell.edu/comet/cover_haser.html}} given that our aperture does not contain the full flux for the species of interest, we compute $Q_{\rm C_{\rm 2}}$ of (1.1 $\pm$ 0.2) $\times$ 10$^{\rm 26}$ molec s$^{\rm -1}$ and $Q_{\rm CN}$ of (0.7 $\pm$ 0.1) $\times$ 10$^{\rm 26}$ molec s$^{\rm -1}$. The resultant $Q$ ratio $\log$[$Q_{\rm C_{\rm 2}}$/$Q_{\rm CN}$] is $\sim$0.19. Our $Q_{\rm C_{\rm 2}}$ and $Q_{\rm CN}$ estimates are consistent within 2$\sigma$ and a factor two, respectively, with the values measured at 0.25 au farther from the Sun than our observation \citep{Jehin2022}: $Q_{\rm C_{\rm 2}}$ = (1.88 $\pm$ 0.42) $\times$ 10$^{\rm 26}$ molec s$^{\rm -1}$ and $Q_{\rm CN}$ = (2.09 $\pm$ 0.13) $\times$ 10$^{\rm 26}$ molec s$^{\rm -1}$.

The carbon-chain composition of K2 can be compared with canonical classification criteria \citep{A'Hearn1995}. A $Q_{\rm CN}$ in \citet{A'Hearn1995} was measured from CN(0,0) violet band at 3 880 \AA, which is out of the nominal wavelength of MUSE. 
The $Q_{\rm CN}$ in Table \ref{t02} was computed by the flux integrated over 9 141--9 269 $\AA$, which is narrower than its nominal band region of $\sim$9 100--9 320 $\AA$, thereby missing part of the CN(1,0) signal. If we simply assume a Gaussian band shape of CN(1,0) over the nominal wavelength, the approximate missing fraction is $\sim$25 \% of the total CN(1,0) band signal. Hence, considering the CN violet band that is generally stronger than the CN(1,0) red band \citep{Schleicher2010}, resulting $Q_{\rm CN}$ of the CN violet band with a flux (a few times brighter than our $F_{\rm CN}$) and corresponding g-factor ($\sim$3--4 times the CN(1,0) value; \citealt{A'Hearn1995}) may be comparable to our $Q_{\rm CN}$ in the end.
The log ratios of the production rates $\log$[$Q_{\rm C_{\rm 2}}$/$Q_{\rm CN}$] are anyhow well within the average range of Oort-cloud comets \citep{A'Hearn1995,Fink2009,Cochran2012} and in the typical group of comets in carbon-chain composition having $\log$[$Q_{\rm C_{\rm 2}}$/$Q_{\rm CN}$] $\geq$ $-$0.18 \citep{A'Hearn1995}. By comparison, the $\log$[$Q_{\rm C_{\rm 2}}$/$Q_{\rm CN}$] value reported for K2 one month before the time of our observation is $\sim$ $-$0.04 \citep{Jehin2022}. This value is still within the typical carbon-chain group but lower (i.e., carbon deficient) than our estimate, suggesting that the C$_{\rm 2}$/CN ratio may change with the distance from the Sun.

\section{Discussion\label{sec:dis}}

When compared to a handful of comets observed at similar preperihelion distances ($r_{\rm H}$ = 2--3 au), K2's Q$_{\rm C_{\rm 2}}$ and Q$_{\rm CN}$ are in line with those of active Oort-cloud comets \citep{A'Hearn1995,Fink2009}, and overall more prominent relative to short-period comets and interstellar comet 2I/Borisov by up to three orders of magnitude (e.g. \citealt{Fink2009,Cochran2012,Snodgrass2016,Opitom2019,Bannister2020}). However, while some of those comets in abundant carbon-chain molecules also display a significant level of NH$_{\rm 2}$ emissions, K2 lacks relevant spectral signals (Fig. \ref{Fig01}). 
This apparent deficiency in NH$_{\rm 2}$ is either due to a real depletion of the molecule or the temporal status, given that NH$_{\rm 3}$, the possible parent molecule of NH$_{\rm 2}$, can increase its production rate at a smaller heliocentric distance (e.g. \citealt{DelloRusso2016}).

Although our single-epoch dataset may hardly provide a clear picture of comet composition, it is worthwhile to compare K2's characteristics (carbon-chain-rich and NH$_{\rm 2}$-deficient) with another CO-rich Oort-cloud comet C/2016 R2 (PanSTARRS) \citep{McKay2019} that exhibits deficiencies both in NH$_{\rm 2}$ and carbon-chain molecules in the optical (e.g. \citealt{Cochran2018}). 
If the NH$_{\rm 2}$ deficiency of the two CO-rich comets reflects their intrinsic depletion, a lack of the availability of NH$_{\rm 3}$ in the accretion stage requires their place of origin to be at least $\gtrsim$5--10 au \citep{Lodders2004,Kurokawa2020} and to be $>$35 au from their CO richness \citep{Womack2017}.
\citet{Lisse2022} suggest that the CO-richness today presumably pertains to the ejection time of small bodies into the Oort Cloud, as earlier ejection ensures their nuclei retain inherent CO ices. In this regard, the observed difference in photo-dissociated gas molecules between the two comets that are likely to have shared similar temperature ranges may support the idea that the compositional heterogeneity in the optical is highly individual \citep{A'Hearn1995,Cochran2012,Opitom2019}.

The asymmetric coma environment around the water ice sublimation boundary suggests that K2's nucleus surface may not be uniformly active, resulting in three distinct dust populations inhabiting different parts of the coma (Sect. \ref{sec:res2}): near-nucleus chunks, steady-state outer envelope, and anti-sunward jet. 
Only C$_{\rm 2}$(0,0) aligns with the dust envelope among the three gases detected, while the other species (OI$^{\rm 1}$D, and CN(1,0) red) appear co-spatial with the near-nucleus continuum enhancement (Sect. \ref{sec:res3}).
The possible association between the dust chunks and OI$^{\rm 1}$D indicates that significant amounts of H$_{\rm 2}$O ice may be present in the dust. In conjunction with their colocation in the coma region where dust color is redder than the average (Fig. \ref{Fig02}b), these connections propose that the scattering medium could be dirty ice (refractory fraction of $\gtrsim$1 \%; \citealt{Mukai1986}) whose size sufficiently exceeds the wavelength of observation \citep{Hadamcik2009,Kwon2022a}. This idea of large-sized dust containing H$_{\rm 2}$O ice is in line with a recent report of K2 in its near-infrared spectroscopic observations that display evidence of the characteristic H$_2$O absorption bands at 1.5 and 2.0 $\mu$m and suggest non-zero dust refractory contents in the ice particles \citep{Protopapa2022}. In our dynamical modeling of coma dust (Fig. \ref{Fig03}c), only dust particles with low mobility ($\beta$ $<$ 0.001, $>$0.1 mm; \citealt{Kwon2017}) emitted long ago ($r_{\rm H}$ $>$ 4.1 au, released at least six months before the observation) can match the small radial scale ($\rho$ $\lesssim$ 10$^{\rm 3}$ km) of the dust clustering.
This mm-sized dust is in accordance with the size range predicted from distant cometary activity \citep{Bouziani2022} and with previous estimates of K2 at large heliocentric distances \citep{Jewitt2017,Hui2017}.



The ejection of dust containing H$_{\rm 2}$O ice necessitates the sublimation of ices more volatile than H$_{\rm 2}$O, such as CO$_{\rm 2}$ or CO (e.g. \citealt{Gundlach2020,Ciarniello2022}). K2's unprecedentedly high coma activity beyond Saturn \citep{Meech2017,Jewitt2017} and direct detection of CO emission at $r_{\rm H}$ $\sim$ 6.72 au \citep{Yang2021} underpin the presence of abundant supervolatile ice in the nucleus of the comet. It is worth noting that the CO emission in K2 was headed primarily in the sunward direction \citep{Yang2021}, the same direction where dust chunks are dominating in the binned image (Fig. \ref{Fig02}b). 
Having the H$_{\rm 2}$O ice-containing chunks may offer a plausible answer to the observed discrepancy of K2 between optical and infrared wavelengths (Sect. \ref{sec:intro}). Around the time of our observations, the comet is still far away from the Sun ($\sim$2.5--2.7 au). Given that the lifetime of ice decreases as the temperature increases, the low temperature may hamper direct H$_{\rm 2}$O-ice sublimation from the nucleus. Dust of $\gtrsim$100 $\mu$m in size has sublimation lifetime of $\sim$10$^{\rm 4-5}$ sec (corresponding to $\gtrsim$10$^{\rm 4}$ km for the dust) at $\sim$2.5 au \citep{Mukai1986}. The slit width of ESO/CRIRES$+$ that detected no H$_{\rm 2}$O emission features of small ice grains samples about 1 500 km from the nucleus, that is, at distances shorter than those where sublimation takes place (Lippi et al. in prep.). On the other hand, our spectroscopic measurements and those reported at similar epochs (e.g. \citealt{Jehin2022} using 0.6-m TRAPPIST robotic telescopes) have been carried out over $\gtrsim$10$^{\rm 4}$ km from the nucleus, thereby having a much higher chance to detect the emission features of daughter molecules produced from the photo-dissociation process. 
In this context, the detection of absorption features of H$_{\rm 2}$O ice in K2 by IRTF InfraRed Telescope Facility (IRTF)/SpeX \citep{Protopapa2022} provides supportive evidence of the presence of dust chunks ($\gtrsim$100 $\mu$m; \citealt{Grundy1998}). 

Taken altogether, the K2 coma environment investigated in this study can be interpreted as the following evolutionary scenario: dirty ice chunks entrained by supervolatile ice at very large heliocentric distances have started sublimating its internal H$_{\rm 2}$O ice as the comet reaches the inner solar system while an anti-sunward jet has formed near the water ice sublimation boundary, presumably launched by H$_{\rm 2}$O ice. Monitoring observations of this active comet during the crossing of this normal water-ice sublimation boundary were conducted using optical imaging polarimetry (Kwon et al. in prep.) and infrared echelle spectroscopy (Lippi et al. in prep.). For this reason, we defer to discussing K2's secular evolution in a broader context for our future research.
\\

\section{Summary \label{sec:sum}}

Here we present a new IFU observation of K2 around the water ice sublimation boundary using VLT/MUSE, as part of our monitoring observation program, in conjunction with  ZTF archival data. The coma appears heterogeneous, consisting of three distinct populations of dust (near-nucleus chunks, envelope, and jet distributed in sizes, colors, and possibly compositions) and three significant gas (OI$^{\rm 1}$D and C$_{\rm 2}$(0,0) and CN(1,0) red bands) occupying different parts of the coma. Based on their spectral and spatial information, we discuss dust properties, abundance and production rates of the gas, and their mutual association. 

If water ice-harboring large dust chunks were indeed present near the nucleus of K2, this would be compatible with the existence of water-ice grains likely containing nonzero refractory content as suggested by infrared spectroscopic observations taken at similar epochs \citep{Protopapa2022}, further explaining its bare detection of dust continuum and H$_{\rm 2}$O emission features (\citealt{Lippi2022}, Lippi et al. in prep.).
All physical quantities derived from this study can vary as K2 travels through the inner solar system
(e.g. NH$_{\rm 2}$ deficiency and Q$_{\rm C_{\rm 2}}$/Q$_{\rm CN}$; \citealt{Cochran2012,Bannister2020}). 
We thus expect this study to serve as a reference of cometary activity around the water ice sublimation boundary for future studies that track the secular evolution of cometary activity.
\\

\begin{acknowledgements}

Based on observations made with ESO Telescopes at the Paranal Observatory under program 109.24F3.001.
Y.G.K gratefully acknowledges funding from the Volkswagen Foundation. M.L. acknowledges funding from the European Union’s Horizon 2020 research and innovation program under grant agreement no. 75390 CAstRA.

\end{acknowledgements}


\begin{appendix}
\section{Ancillary figures \label{sec:app1}}

\begin{figure}[!b]
\vskip-1ex
\centering
\includegraphics[width=8.9cm]{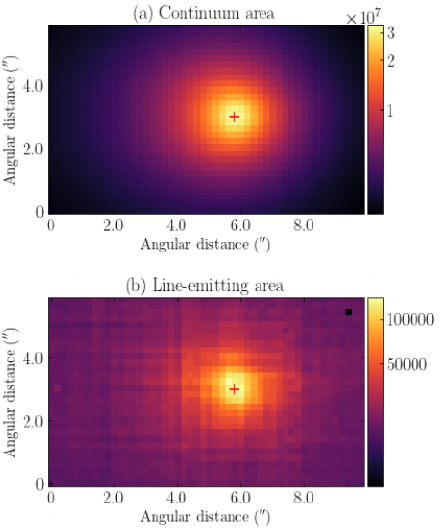}
\caption{Unbinned flux image of (a) the continuum-emitting area and (b) the line-emitting area with contributions from all gas species. Units of axes and color bars are pixels and flux in 10$^{\rm -20}$ erg cm$^{\rm -2}$ s$^{\rm -1}$, respectively. One pixel corresponds to $\rho$ $\sim$ 270 km, such that the cropped FoV covers the $\sim$8 100 km $\times$ 13 500 km rectangular coma region. The red plus ($+$) marks the optocenter. The images are aligned as in Figure \ref{Fig02}.}
\label{Figap01}
\vskip-1ex
\end{figure}

Figure \ref{Figap01} displays unbinned continuum and gas emission flux images acquired from the continuum subtraction. Both coma components have a strong concentration in the central part of the coma without significant large-scale asymmetry. The maps are constructed by combining the two cubes that have the smallest seeing (0\farcs39 and 0\farcs5) in our datasets. The compact distribution appears rather concentric around the optocenter than a sunward extension shown in the binned images (Figs. \ref{Fig02}b and \ref{Fig04}). 
Despite the binned images having less influence from the seeing effect, we cannot rule out external factors such as the centering effect of the telescope that could have led to the apparently preferred alignment.
As such, we refrain from making any definitive statements regarding the near-nucleus morphology other than the strong central condensation, following our criteria (Footnote \#4 on page 2).

\begin{figure*}[!b]
\vskip-1ex
\centering
\includegraphics[width=\textwidth]{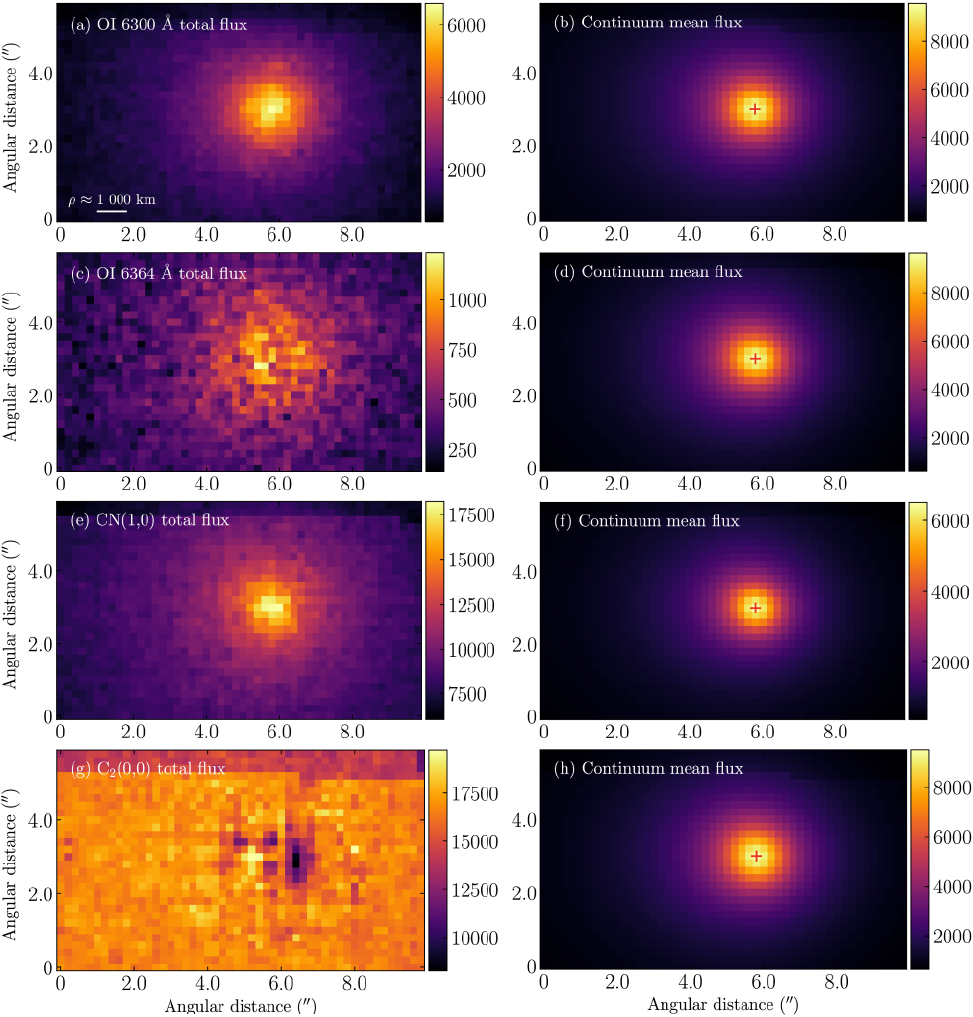}
\caption{{\it Left}: Unbinned flux distribution of each gas component highlighted in Figure \ref{Fig04}b. All fluxes are in units of 10$^{\rm -20}$ erg cm$^{\rm -2}$ s$^{\rm -1}$.  The flux drop of CN(1,0) red and C$_{\rm 2}$(0,0) on the top part (panels e and g) is an artifact likely due in part to their flux in the outer coma region highly susceptible from background noise and thus having large uncertainties in estimating the continuum signal.
 {\it Right}: Unbinned flux distribution of the underlying continuum integrated over the considered wavelengths. They are subtracted from the observed spectrum to offer the line flux on the left side. The red plus ($+$) marks the optocenter. }
\label{Figap03}
\vskip-1ex
\end{figure*}

Ancillary information of the spectral analysis is available in Figure \ref{Figap03}. The figure summarizes unbinned flux distributions of the gas species highlighted in Figure \ref{Fig04}b in the left column and the underlying dust continuum within the integrated wavelength interval in the right column. Both emission and continuum signals are consistent with the trends that appeared in the binned images, where most of the light from OI$^{\rm 1}$D 6 300 \AA\ line (including its OI$^{\rm 1}$D doublet companion at 6 364 \AA) and CN(1,0) red band come from the near-nucleus region ($\rho$ $\lesssim$ 10$^{\rm 3}$ km) while C$_{\rm 2}$(0,0) Swan band peaks off from the optocenter by $\sim$500--1 000 km.
This is also compatible with the trend that the inner coma ($\rho$ $\approx$ 1 000 km) contains negligible emission features of C$_{\rm 2}$(0,0) Swan band relative to the outer coma ($\rho$ $\approx$ 20 000 km; Fig. \ref{Fig01}a). The continuum fluxes underlying each gas component distributes nearly the same, forming a compact spherical coma around the optocenter.

\begin{figure*}[!b]
\vskip-1ex
\centering
\includegraphics[width=\textwidth]{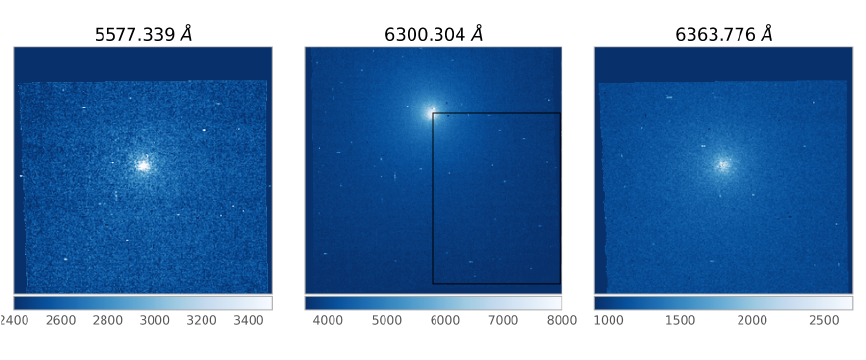}
\caption{Maps of the three forbidden oxygen lines. The FoV is 60 $\times$ 58 \arcsec$^{\rm 2}$, and the images are oriented North up and East left. The 6 300 $\AA$ line map is offset with respect to the other two maps since the spatial extension of the emission is larger and only off-centered observations were used. The black box illustrates the aperture used for the measurement of the water production rate.}
\label{Figap04}
\vskip-1ex
\end{figure*}

\end{appendix}

\end{document}